\documentstyle[aps,prd,floats,epsfig,url]{revtex}

\newcommand{\vek}[1]{{\mbox {\boldmath $#1$}}}

\newcommand{\href}[2]{#1}

\begin{document}             


\title{Neutralino Gamma-ray Signals from Accreting Halo Dark Matter}

\author{
Lars Bergstr\"om\footnote{E-mail: lbe@physto.se},
Joakim Edsj\"o\footnote{E-mail: edsjo@physto.se} and
Christofer Gunnarsson\footnote{E-mail: cg@physto.se}}

\address{Department of Physics, Stockholm University, Box 6730, 
SE-113 85 Stockholm, Sweden}

\maketitle    

\begin{abstract}

There is mounting evidence that a self-consistent model for particle
cold dark matter has to take into consideration spatial
inhomogeneities on sub-galactic scales seen, for instance, in
high-resolution $N$-body simulations of structure formation.  Also in
more idealized, analytic models, there appear density enhancements in
certain regions of the halo.  We use the results from a recent
$N$-body simulation of the Milky Way halo and investigate the
gamma-ray flux which would be produced when a specific dark matter
candidate, the neutralino, annihilates in regions of enhanced density. 
The clumpiness found on all scales in the simulation results in very
strong gamma-ray signals which seem to already rule out some regions
of the supersymmetric parameter space, and would be further probed by
upcoming experiments, such as the GLAST gamma-ray satellite.  As an
orthogonal model of structure formation, we also consider Sikivie's
simple infall model of dark matter which predicts that there should
exist continuous regions of enhanced density, caustic rings, in the
dark matter halo of the Milky Way.  We find, however, that the
gamma-ray signal from caustic rings is generally too small to be
detectable.

\end{abstract} 

\section{Introduction}
\label{sec:intro}

Recent determinations of cosmological parameters have singled out a
region of the matter density $\Omega_M\sim 0.3$ clearly larger than
allowed by big bang nucleosynthesis.  This coupled with many other pieces of
evidence makes the existence of non-baryonic dark matter compelling
\cite{lbreview}.  However, we still have no clue as to the nature of
the dark matter other than that it plausibly exists in the form of
non-relativistic (cold) particles.  If the particle is massive and has
weak-interaction coupling to ordinary matter, i.e., it is a WIMP (weakly
interacting massive particle), there are good prospects for its
eventual experimental detection.  The lightest supersymmetric
particle, usually a neutralino, is one of the prime candidates.  To
detect or rule out particle dark matter such as the neutralino is
obviously an important experimental undertaking.  However, most
detection methods depend quite sensitively not only on the properties
(the exact values of the mass and cross sections) of the candidate
particle itself, but also on the distribution of dark matter in our
Galactic halo.  This is starting to be probed in computer simulations
and to some extent also through analytical modelling of the formation
history of dark matter halos.

The currently most fashionable model of structure formation is that of
primordial fluctuation-seeded hierarchical clustering, where $N$-body
simulations are beginning to have high enough resolution to give
information on sub-galactic scales \cite{nbody}.  In this class of
models, galactic halos usually have a very complicated merging history
leading, in the infall picture, to extensive irregular foldings of the
initially thin phase sheets on which the dark matter particles were
lying at the time of kinetic decoupling from the primordial plasma. 
As shown in a recent work by Calc\'aneo-Rold\'an and Moore
\cite{moore}, the Galactic halo in this scenario contains a lot of
substructure leading to significant possible enhancements of the
annihilation rate in the overdense regions.  Since the annihilation
rate is proportional to the square of the WIMP density, the gamma-ray
signal in the direction of these galactic halo clumps should be
considerably enhanced compared to the case of a smooth halo profile,
which has most frequently been considered in previous analyses.

In another, highly idealized model, having the virtue of being
analytically treatable, proposed by Sikivie
\cite{sikivie981,sikivie982,sikivie992}, continuous infall of dark
matter on our galaxy should give rise to ring shaped caustics of dark
matter.  If the velocity dispersion of the infalling particles is
sufficiently small, the caustics could contain significant
overdensities, again with a possible detectable gamma-ray flux as a
result.

Some general results on the increased indirect detection signals of
supersymmetric dark matter in a clumpy halo were obtained in
\cite{clumpy}.  With the models mentioned we now have two specific
scenarios with which to make more quantitative estimates of the
possible enhancements.  In this paper we first investigate the
magnitude of the enhancements from the hierarchical clustering
model.  We adopt the results from the numerical simulations performed
in \cite{moore}, but supplement that analysis with actual values for
the annihilation cross sections which we compute.  We focus on the
neutralino, since, as mentioned, it arises naturally in supersymmetric
extensions of the standard model as a good dark matter candidate, but
our results should be applicable to the more general class of WIMPs.
We are primarily interested
in the gamma ray flux (both continuous and monochromatic lines) since
this is not smeared by propagation uncertainties.
Then we also consider the flux of gamma rays (in this case mostly the
continuous gamma rays) from annihilations in the closest caustic ring
in the model of Sikivie.  This has not been studied before, unlike the
case of direct detection, where the caustic flows have been shown to
lead to some interesting possible effects, such as a reversal of the
annual modulation pattern caused by the motion of the Solar system in
the halo \cite{krauss}.

We work in the Minimal Supersymmetric Standard Model (MSSM) (see
\cite{lbreview,jkg} for reviews of supersymmetric dark matter).
We also estimate the increased flux of antiprotons which is correlated
to the continuous gamma ray flux and compare it to the BESS 97
measurements \cite{bess} on antiprotons. 
 
In the next Section we briefly review the signal patterns and fluxes
expected for a given halo model.  In Section~\ref{sec:neugamma} we
will define the MSSM framework we work in and describe how the gamma
ray yield is calculated for a given MSSM model.  In
Section~\ref{sec:nbody} we compute the gamma-ray flux in the
hierarchical clustering model, then the Sikivie model for caustic
rings and its implications is treated extensively in
Section~\ref{sec:caustics}.  Finally, we conclude in
Section~\ref{sec:conclusions}.

\section{Gamma-ray Signals - general considerations}

\subsection{Signal fluxes}

Substructure in the Galactic halo may be weak and narrow features on
the sky so a telescope with large detection area and good angular
resolution might be preferable to a telescope with small area and
large angular acceptance, $\Delta\Omega$.  However, any search for
halo features has to face the uncertainty in the location of these
narrow features which thus may be difficult to find.

Perhaps the best strategy would be to use a large angular acceptance
detector like the GLAST satellite \cite{glast} to search for extended
structures such as ``hot spots'' in the gamma-ray sky or the ring-like
pattern expected from the caustic rings, and once discovered, their
detailed properties could be investigated with a telescope of larger
area but smaller angular acceptance, like the Air Cherenkov Telescopes
(ACTs) currently being planned or built \cite{act}.  As an aside, it
may be mentioned that in the EGRET catalog of unidentified point
sources with steady emission, there could in principle be a
contribution from the ``exotic'' gamma-ray sources discussed here.

The $\gamma$-ray flux from WIMP annihilations in the galactic halo is
given by \cite{bergstromullio}
\begin{equation}
  \label{eq:flux1}
  \Phi_{\gamma}(\eta)=\frac{N_{\gamma}\sigma v}{4\pi
  m_{\chi}^{2}}\int_{L}{\cal D}^{2}(\ell )\ d\ell (\eta),
\end{equation}
where ${\cal D}(\ell)$ is the halo mass density of WIMPs at distance
$\ell$ along the line of sight.  We will focus on the gamma ray flux
off the galactic plane, and define $\eta$ to be the angle between the
direction of the galactic center and the line of sight in in a plane
perpendicular to the galactic disk (and with both the Earth and the
galactic center in the plane).  $\eta$ is thus equivalent to the
galactic latitude, except that it can take on values larger than
$90^{\circ}$, with $\eta=180^{\circ}$ corresponding to the
anti-galactic center.  We assume that the Earth is located in the
$z=0$ plane.  The integral is carried out along the line of sight,
$L$.  $N_{\gamma}$ is the number of photons created per annihilation. 
In the case of continuous gamma rays, we will compute the integrated
flux above 1 GeV, so $N_{\gamma}$ is the number of photons above 1 GeV
per annihilation and $\sigma v$ is the total annihilation cross
section times the relative velocity of the annihilating particles,
i.e., the annihilation rate.  We will also give predictions for the
annihilation into the final states $\gamma\gamma$ and $Z\gamma$ which
give monochromatic photons; in this case $N_\gamma$ is 2 and 1,
respectively.  (Of course, the $Z$ boson in the $Z\gamma$ final state
will also give gamma-rays in its decay, but these mainly populate low
energies and are included in the continuous gamma ray flux.)  To
obtain the flux for a specific angular acceptance we also have to
integrate over $\Delta\Omega$.

To factorize the part that depends on the particle physics model from
the part that depends on the halo structure, we can write the gamma
ray flux as
\begin{equation}
  \label{eq:split}
  \Phi_{\gamma}(\eta;\Delta\Omega)={\cal S}\cdot J(\eta;\Delta\Omega),
\end{equation}
where the particle physics dependent part is
\begin{equation}
  \label{eq:S}
  {\cal S}=\frac{N_{\gamma}\sigma v}{m_{\chi}^{2}}
\end{equation}
and the halo structure-dependent part is
\begin{equation}
  \label{eq:jofeta}
  J(\eta; \Delta\Omega) =
  \frac{1}{4 \pi}\int_{\Delta\Omega}\!\int_{L}{\cal D}^{2}
  (\ell)\ d\ell (\eta)d\Omega.
\end{equation}
We will in the following also use the solid angle
average of $J(\eta)$:
\begin{equation}
    F_{\Delta\Omega}(\eta) = \frac{J(\eta; \Delta\Omega)}{\Delta\Omega}.
\end{equation}

\subsection{Background estimates}

The diffuse $\gamma$-ray background has been measured by EGRET
\cite{egret} and can be approximately fit \cite{piero} by 
\begin{equation}
  \label{eq:fit}
  \frac{dN(E_{\gamma},l,b)}{dE_{\gamma}} = 
  N_0(l,b)\left( \frac{E_{\gamma}}{1\ {\rm 
  GeV}}\right)^{\delta} \
  10^{-6}\ 
  {\rm cm}^{-2}{\rm s}^{-1}{\rm GeV}^{-1}{\rm sr}^{-1},
\end{equation}
where 
\begin{equation}
  \label{eq:nnaught}
  N_0(l,b)=\left\{ \begin{array}{ll}
  \protect{\frac{85.5}{\sqrt{1+(l/35)^2}\sqrt{1+(b/1.1+|l|0.022)^2}}}+0.5 & 
  |l|\geq 30^{\circ}\\[2ex]
  \protect{85.5\over{\sqrt{1+(l/35)^2}\sqrt{1+(b/1.8)^2}}}+0.5 & 
  |l|\leq 30^{\circ}\ \\
  \end{array}
  \right.,
\end{equation}
and $l$ and $b$ are the longitude and latitude respectively, in the
sky.  We adopt $\delta=-2.7$ as in Ref.~\cite{piero}.  Since this
parameterization is discontinuous at $b=90^{\circ}$ we smooth it to
join the branches for $b\rightarrow 90^{\circ}, l=0^{\circ}$ and
$b\rightarrow90^{\circ}, l=180^{\circ}$.

\section{Neutralino annihilation as a $\gamma$-ray source}
\label{sec:neugamma}

\subsection{Definition of the MSSM and the neutralino}
\label{sec:mssm}

\begin{table}
  \caption{The ranges of parameter values used in our 
  scans of the MSSM parameter space. Special scans aimed at 
  interesting subregions of this parameter space have also been performed.}
  \label{tab:scans}
  \begin{tabular}{lrrrrrrr}
  Parameter & $\mu$ & $M_{2}$ & $\tan \beta$ & $m_{A}$ & $m_{0}$ & 
  $A_{b}$ & $A_{t}$ \\
  Unit & GeV & GeV & 1 & GeV & GeV & $m_{0}$ & $m_{0}$ \\ \hline
  Min & -50000 & -50000 &  1.0 &     0 &   100 & -3 & -3 \\
  Max &  50000 &  50000 & 60.0 & 10000 & 30000 &  3 &  3 \\
  \end{tabular}
\end{table}

To make specific predictions of the expected gamma-ray fluxes
possible from WIMP annihilation, we will now assume that the
dark matter particle is a supersymmetric, electrically neutral 
particle. We will work in the Minimal Supersymmetric
Standard Model, MSSM \cite{haberkane,jkg}, using the computer
code {\sffamily DarkSUSY} \cite{darksusy} to make our quantitative predictions.
The lightest stable supersymmetric particle is in 
most models the neutralino, which is a superposition of the 
superpartners of the gauge and Higgs fields,
\begin{equation}
  \tilde{\chi}^0_1 = 
  N_{11} \tilde{B} + N_{12} \tilde{W}^3 + 
  N_{13} \tilde{H}^0_1 + N_{14} \tilde{H}^0_2.
\end{equation}
For the masses of the neutralinos and charginos we use the one-loop 
corrections as given in \cite{neuloop} and for the Higgs boson
masses we use the leading log two-loop radiative corrections,
calculated within the Feynman diagrammatic approach with the computer 
code FeynHiggsFast \cite{feynhiggsfast}.

The MSSM has many free parameters, but following common praxis we 
introduce a number of simplifying assumptions which leaves us 
with 7 parameters, which we vary between generous bounds.  The ranges
for the parameters are shown in Table~\ref{tab:scans}.  In total we have
generated about 93\,000 models that are not excluded by accelerator searches. 

We check each model to see if it is excluded by the most recent
accelerator constraints, of which the most important ones are the LEP
bounds \cite{lepbounds} on the lightest chargino mass,
\begin{equation}
  m_{\chi_{1}^{+}} > \left\{ \begin{array}{lcl}
  91 {\rm ~GeV} & \quad , \quad & | m_{\chi_{1}^{+}} -
  m_{\chi^{0}_{1}} | 
  > 4 {\rm ~GeV} \\
  85 {\rm ~GeV} & \quad , \quad & {\rm otherwise}
  \end{array} \right.
\end{equation}
and on the lightest Higgs boson mass $m_{H_{2}^{0}}$ (which range from
91.4--107.7 GeV depending on $\sin (\beta-\alpha)$ with $\alpha$ being
the Higgs mixing angle) and the constraints from $b \to s \gamma$
\cite{cleo}.

We only consider those MSSM models where the neutralinos can make up most of 
the dark matter in our galaxy and therefore impose the cosmological 
constraint $0.05 < \Omega_{\chi}h^{2} <0.5$ where we have calculated 
the relic density according to the procedure described in 
Ref.\ \cite{coann}. Here $h$ is the scaled
Hubble constant, $H_0=h\cdot 100$ km$\,$s$^{-1}$Mpc$^{-1}$, with 
observations giving $h\simeq 0.65\pm 0.15$.

\subsection{Gamma rays from neutralino annihilation}

\begin{figure}
\centerline{\epsfig{file=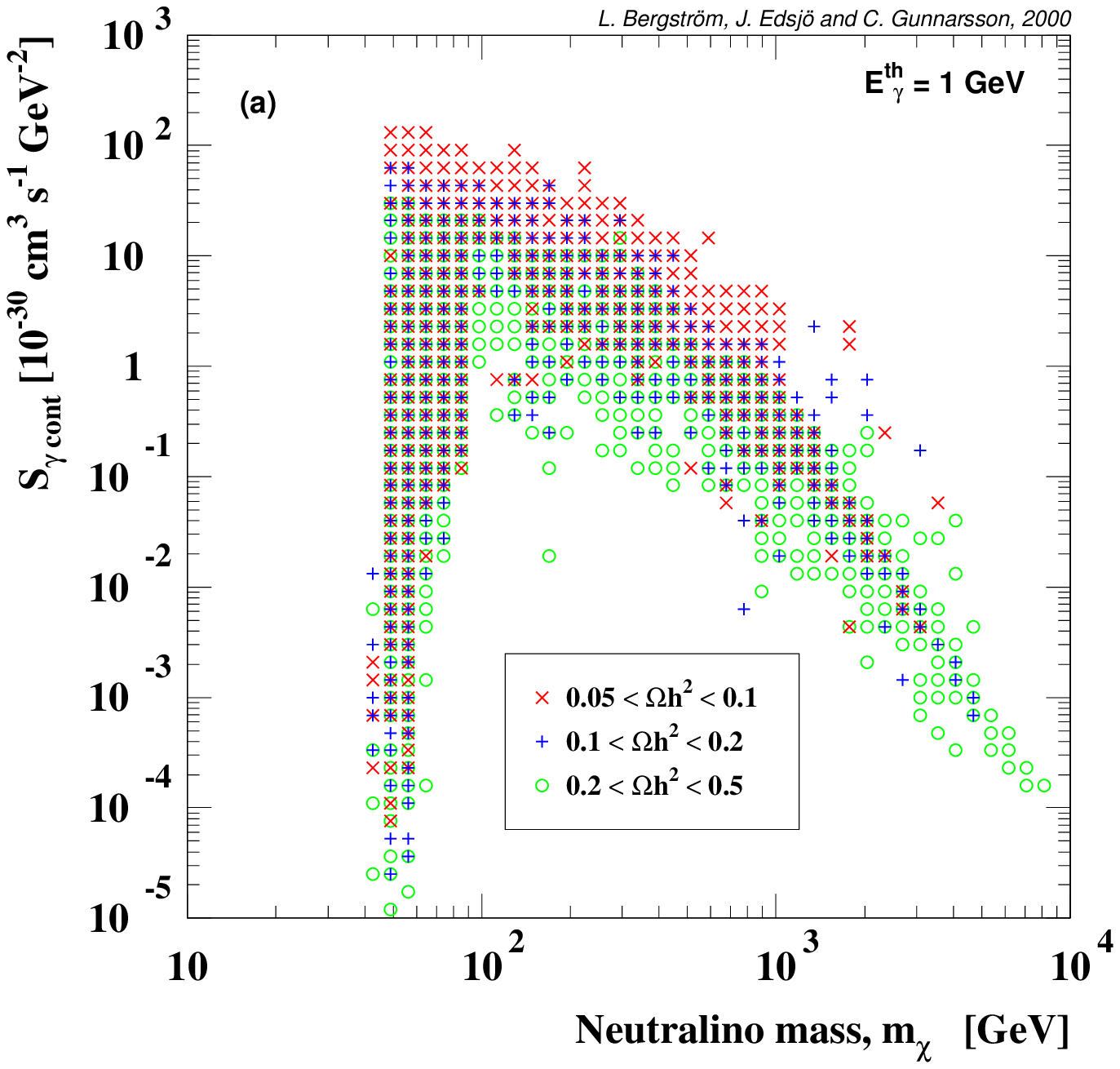,width=0.49\textwidth}
\epsfig{file=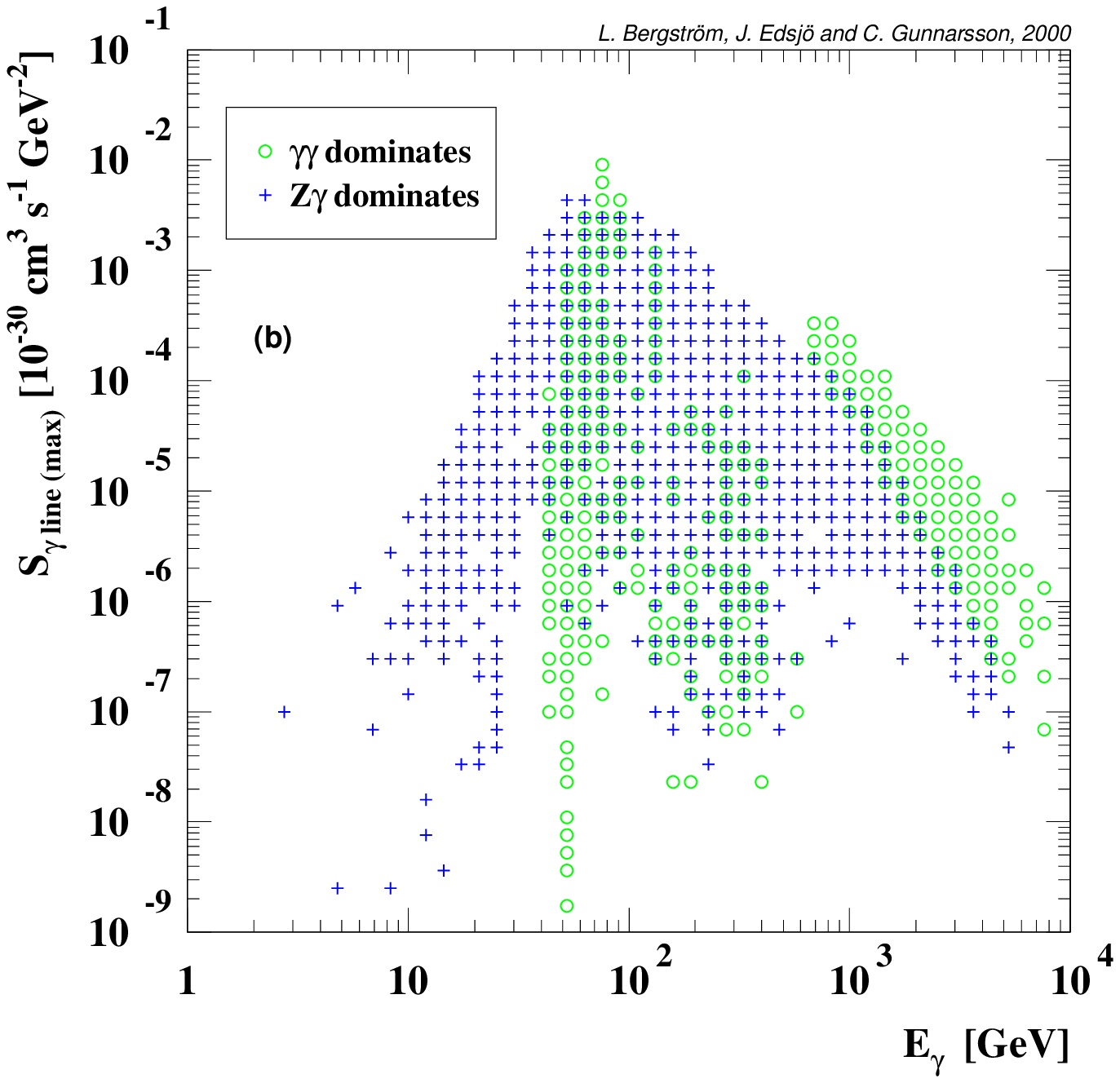,width=0.49\textwidth}}
\caption{
   The ${\cal S}$-factor for gamma rays.  In a) the continuous
   $\gamma$-ray flux above 1 GeV is shown versus the neutralino mass
   and in b) $\max({\cal S}_{\gamma \gamma},{\cal S}_{Z
   \gamma})$ for the monochromatic gamma ray lines is shown versus the
   gamma ray energy.}
\label{fig:sfactors}
\end{figure}    

Gamma rays with a continuous energy spectrum mainly originate from
pions produced in quark jets.  We have simulated the hadronization
and/or decay of the annihilation products with the Lund Monte Carlo
{\sc Pythia} 6.115 \cite{pythia}.  We have also computed the flux of
monochromatic gamma lines that arise from neutralino annihilations to
$\gamma \gamma$ and $Z \gamma$ at the 1-loop level \cite{gammalines},
and which would provide an excellent signature of dark matter if
detected.  In Fig.~\ref{fig:sfactors} we plot the ${\cal S}$-factors
for continuous gamma rays (above 1 GeV) and gamma ray lines
respectively, where ${\cal S}$ is defined in Eq.~(\ref{eq:S}).  The
${\cal S}$-factors have been calculated with {\sffamily DarkSUSY}
\cite{darksusy}.  The maximum for the continuous gamma rays is ${\cal
S}_{\gamma~\!{\rm cont}} \sim 150 \times 10^{-30}$ cm$^{3}$ s$^{-1}$
GeV$^{-2}$ and occurs at $m_{\chi}=57$ GeV whereas the maximum for the
monochromatic gamma ray lines is ${\cal S}_{\gamma\,{\rm line\,(max)}}
\sim 0.0076 \times 10^{-30}$ cm$^{3}$ s$^{-1}$ GeV$^{-2}$, which
occurs for annihilation into $\gamma \gamma$ at $m_{\chi}=78$ GeV. We
will use these maximal values of the ${\cal S}$-factors in our
estimates of the signal below to get the `best-case' scenario with the
highest fluxes.

\subsection{Correlation with antiproton fluxes}

\begin{figure}
\centerline{\epsfig{file=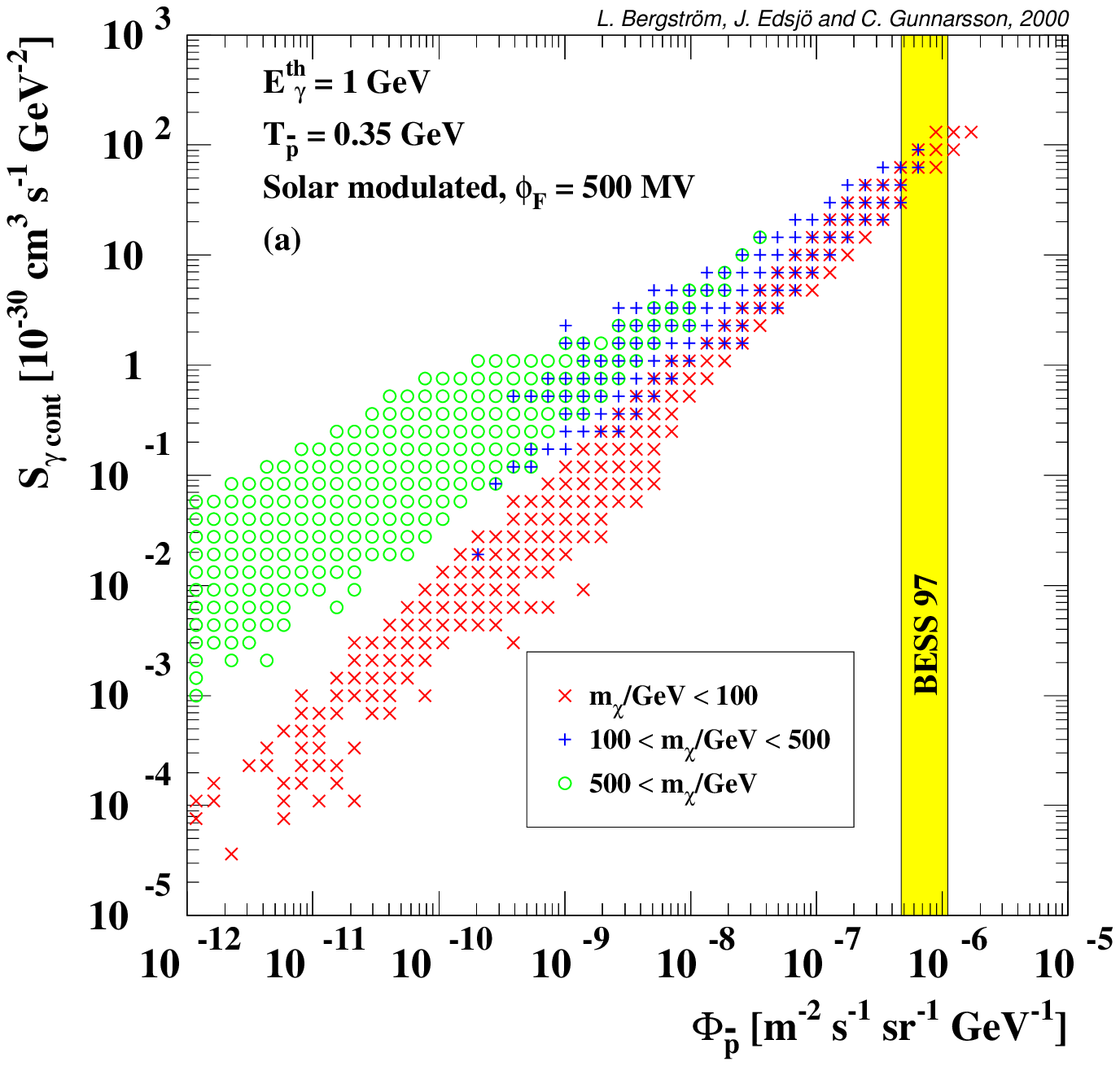,width=0.49\textwidth}
\epsfig{file=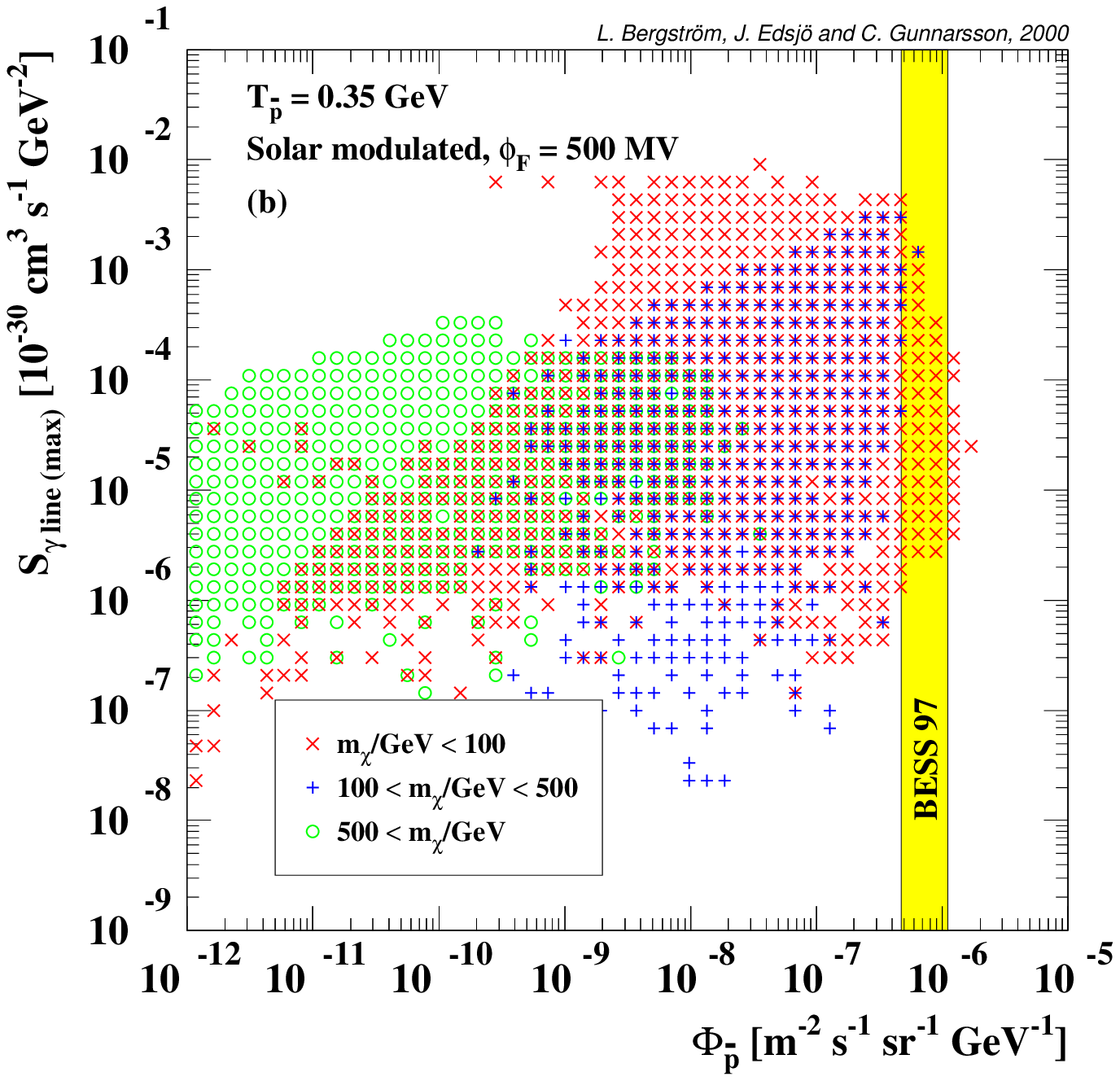,width=0.49\textwidth}}
\caption{The ${\cal S}$ factors for continuous 
$\gamma$s and $\gamma$ lines versus the flux of antiprotons as 
calculated for a smooth halo.}
\label{fig:pbar}
\end{figure}

It is well-known that whenever there is a large annihilation signal in
continuous gamma-rays, there tends to be a large number of antiprotons
also created \cite{clumpy}.  This is due to the fact that both mainly
emanate from quark jets formed in the annihilations.  (On the other
hand, antiprotons and gamma-ray lines are much more weakly correlated
due to completely different production processes.)  Therefore, one has
to check whether the predicted gamma-ray fluxes are consistent with
the present experimental bounds on antiprotons \cite{bess}.

In Fig.~\ref{fig:pbar} we show the ${\cal S}$ factors versus the
antiproton flux as calculated in a smooth halo scenario \cite{pbar}
(with an isothermal sphere halo profile) and as expected, the
correlation between the antiproton flux and the continuous gamma ray
flux is very strong, whereas the correlation with the monochromatic
gamma ray flux is weak.

We will in the following sections give predictions for the gamma-ray
flux in the two structure formation scenarios (hierarchical
clustering or caustics), and we also 
estimate how much the flux of
antiprotons would increase in the two scenarios 
and compare with the BESS bound.

\section{The Hierarchical Clustering Model}
\label{sec:nbody}

\subsection{Results from $N$-body simulations}

We first consider the ``standard'' model of structure formation,
hierarchical clustering of cold dark matter.  Here we make use of the
results in a recent paper by Calc\'aneo-Rold\'an and Moore
\cite{moore}, which we now briefly summarize.  They chose from a large
$N$-body simulation aimed at representing the Local Group, a simulated
dark matter halo at redshift $z=0$ having a peak circular velocity of
around 200 km/s and mass $10^{12}M_\odot$ within the virial radius of
300 kpc.  They computed the local density distribution of this halo by
averaging over the 64 nearest neighbours at the position of each
particle in the simulation.  They then estimated the flux of
annihilation photons by using a discretized version of our
Eq.~(\ref{eq:jofeta}), where the line of sight integral was replaced by
a discrete sum over radial increments of length 1 kpc, and an angular
window size $\Delta\Omega=1^\circ\!\times 1^\circ$ was used for the
binning of a sky map.  Since the halo used showed the characteristic
triaxial, roughly prolate, shape found in $N$-body simulations (with
ratio of short to long axis of 0.5 and intermediate to long axis ratio
of 0.4), it is of non-negligible importance where one puts the
observer (chosen to be 8.5 kpc from the center).  If the long axis is
in the direction of the Galactic center, the flux will obviously be
higher in both the Galactic center and anticentre direction than if
one of the shorter axes is in that direction.  The difference can be
almost an order of magnitude in directions away from the galactic 
center, which is an interesting point to notice,
since it is independent of the existence of substructure. We will use 
the result where the Solar System is put on the short axis.

\begin{figure}
\centerline{\epsfig{file=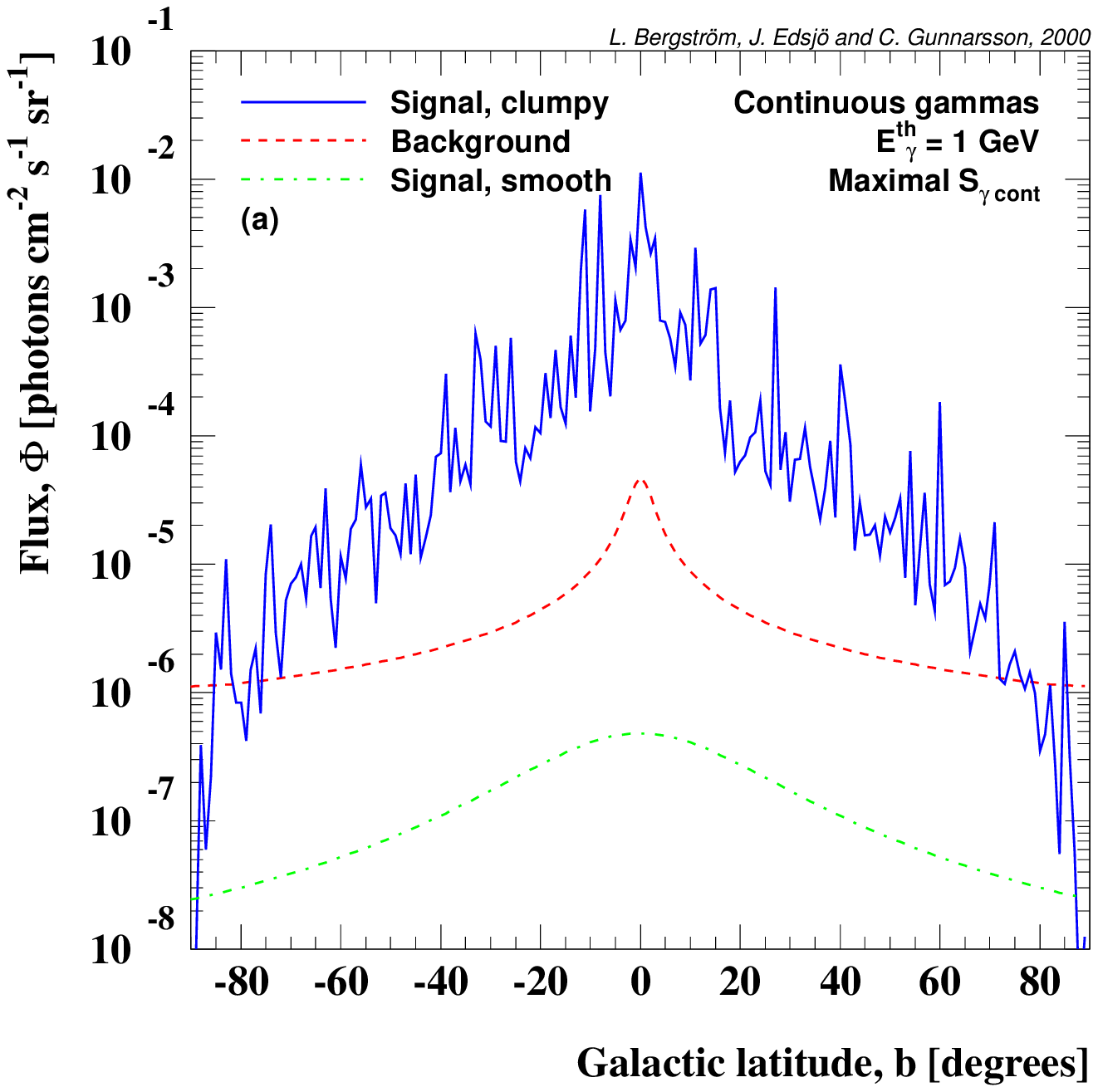,width=0.49\textwidth}
\epsfig{file=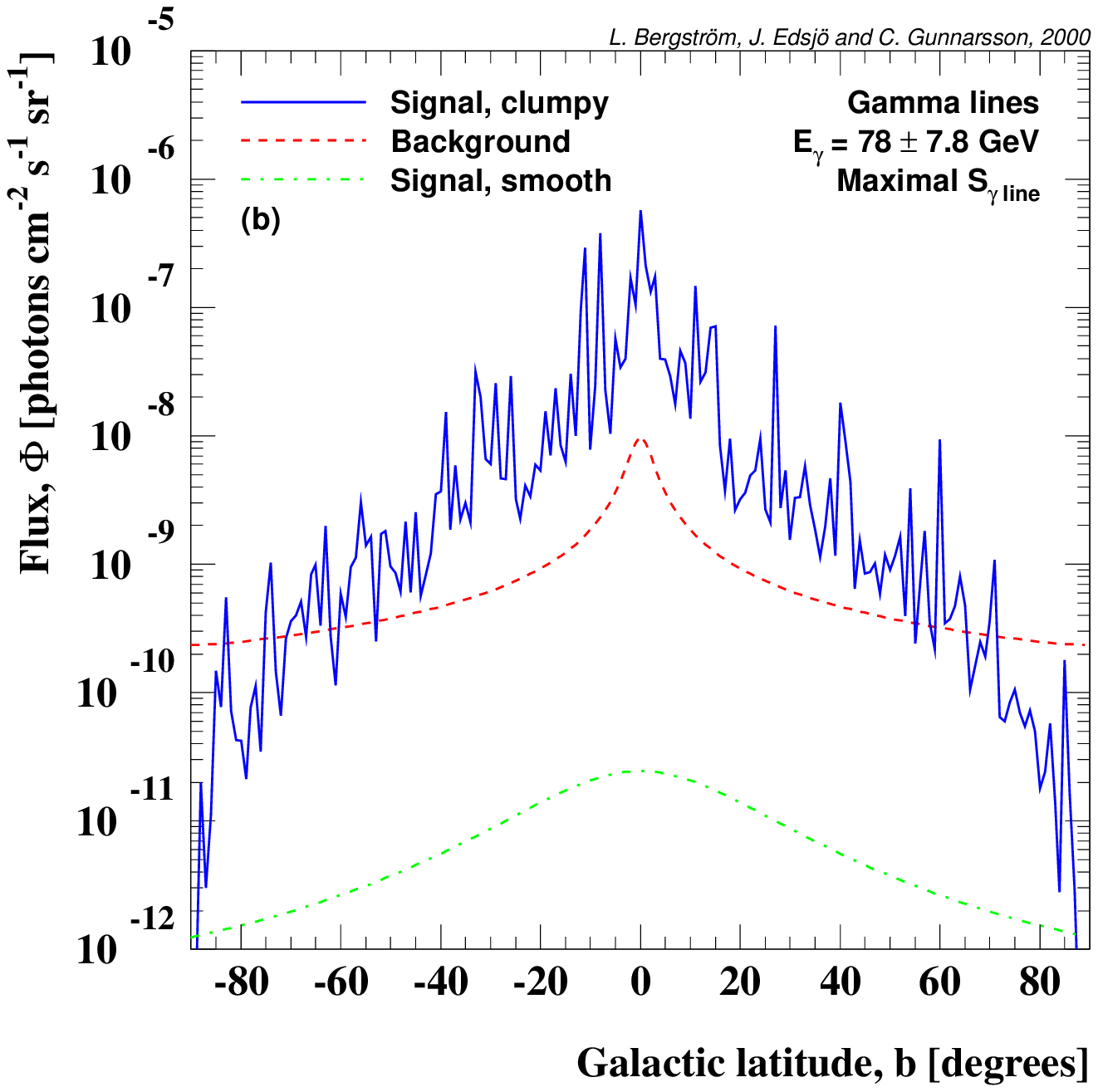,width=0.49\textwidth}}
\caption{Gamma ray fluxes from
substructures as seen in the $N$-body simulations in \protect\cite{moore}. 
Shown is the flux averaged in strips 1$^{\circ}$ high and 44$^{\circ}$ 
wide as a function of the galactic latitude 
(based on Fig.\ 10 in \protect\cite{moore} and on our computed
${\cal S}$-factors in Fig.~\protect\ref{fig:sfactors}). In a) the 
continuous flux above 1 GeV is shown and in b) the flux of monochromatic gamma 
lines is shown. In both figures, the expected diffuse background, 
Eq.~(\ref{eq:fit}) and the expected flux from a smooth isothermal 
sphere are shown for comparison.}
\label{fig:moore}
\end{figure}

In the simulations, substructure seems to be abundant on all scales,
even down to velocity dispersions of a few meters per second, with a
radial profile in the clumps being consistent with a very steep
$\rho\propto r^{-1.5}$ behaviour.  This makes the prediction of the
flux very uncertain, since the line of sight integral will diverge
unless a cutoff is introduced.  A physically unavoidable cutoff will
eventually be set by the self-annihilation rate of the dark matter
particles.  Unfortunately, the mass involved near the cusps of these
substructure clumps is quite small and may be strongly affected by
interaction with the baryonic component.  This interplay of
non-baryonic and baryonic matter is presently very poorly understood,
to the point that even the existence of any dark matter substructure
at all in cold dark matter halos is being disputed. In lack of a good 
description of the interplay of non-baryonic and baryonic matter, the 
only softening of the singularity that is included in the calculation 
of $J(\eta)$ is the self-interaction cut-off. 

To get an estimate of the gamma ray fluxes expected in the
hierarchical clustering scenario we will use Fig.~10 in Ref.\
\cite{moore} where the integral along the line of
sight was calculated, i.e.\ essentially our Eq.~(\ref{eq:jofeta}) 
as a function of
the galactic latitude, $b$ (or $\eta$ in our notation).  
An angular resolution of about $\Delta
\Omega=1^{\circ}\!\times 1^{\circ}$ was assumed and the flux was
averaged over a strip of
height 1$^{\circ}$ and width 44$^{\circ}$.  In Fig.~\ref{fig:moore},
we have plotted the flux expected for the MSSM models giving the
maximal continuum flux and the maximum line flux in the hierarchical
clustering scenario.  The diffuse background also shown can be viewed
as a limit on the unexplained observed gamma ray flux.\footnote{It
should be noted that the background flux between around 30 and 300 GeV
has not been measured but is an extrapolation of EGRET data. 
Only with the upcoming ACTs and, in particular,
with the GLAST satellite will this energy range be measured with
precision.} We have also plotted the flux that we would expect in a
smooth halo scenario, where we have used an isothermal sphere,
see Eq.~(\ref{eq:isodens}) below, with a scale radius of $a_{\rm c}=4$ kpc, 
our galactocentric distance, ${\cal R}_{0}=8.5$ kpc, and the local 
halo density, ${\cal D}_{0}=0.3$ GeV/cm$^{3}$. 
As can be seen, the expected flux in the hierarchical clustering
scenario is very high and the maximal MSSM model chosen here would in
fact already be excluded for this scenario.

It is intriguing that the angular distribution of the diffuse flux
measured by EGRET is consistent with a contribution from neutralino
annihilation giving a peak in the direction of the Galactic center.
However, convincing evidence of a signal can only be obtained 
when GLAST provides also the energy spectrum in the interesting range.
Of course,  detection of a gamma-ray line would be a striking verification
of the WIMP annihilation hypothesis. 

With the fluxes given in Fig.~\ref{fig:moore}, we can estimate the
event rates with GLAST. Let's focus on the peak at a galactic latitude
of $b \simeq -33^{\circ}$.  For continuous gammas, the flux in this
peak is about $7 \times 10^{-4}$ cm$^{-2}$ s$^{-1}$ sr$^{-1}$. 
Assuming an effective area for GLAST of $\langle A_{\rm
eff}\rangle=5000$ cm$^{2}$ and an integration time of 1 year, this
corresponds to $1.5 \times 10^{6}$ events.  Hence, the peak would
easily be visible with GLAST, even after a reduction by a factor of 
ten needed for consistency with the existing background measurements. 
For the gamma ray lines, the fluxes are
lower, and for the same peak at $b \simeq -33^{\circ}$ the flux is
about $3\times 10^{-8}$ cm$^{-2}$ s$^{-1}$ sr$^{-1}$.  This would
correspond to about 60 events in GLAST, which since these photons are
monochromatic would also be easy to see. We also see from the figure 
that there are other peaks with even higher fluxes that would give 
even higher event rates in GLAST.

With Air Cherenkov Telescopes (ACTs), sensitive to gamma radiation,
these signals would also be visible, but since these need to be
pointed in the a priori unknown directions of the overdensities, they
can mainly be used for follow-ups if GLAST would see indications of an
enhanced flux.

\subsection{Antiprotons}

We now estimate the increase of the antiproton flux in the
hierarchical clustering scenario compared to the smooth halo scenario. 
The antiproton flux depends essentially on the average of ${\cal 
D}^{2}$ (with ${\cal D}$ being the neutralino halo density)
within the closest
few kpc.  We do not have access to the full $N$-body simulation
results, but estimate that the increase of the integral of ${\cal
D}^{2}$ locally is about the same as the increase in the gamma ray
flux at high galactic latitudes.  From Fig.~\ref{fig:moore}, we read
off that this increase is about a factor of 5--10 compared to the 
smooth halo scenario.  Hence, we expect
that the antiproton fluxes in Fig.~\ref{fig:pbar} 
would increase by roughly this factor. 
For the gamma ray lines, where the correlation between the gamma ray
and antiproton signal is weak, we can easily find MSSM models with
high gamma ray fluxes that wouldn't violate the BESS bound on
antiprotons.  For the continuous gamma ray fluxes, where the
correlation is stronger, the MSSM models with the highest gamma ray
fluxes, would produce an antiproton flux that is a factor of 5--10
higher than the current BESS bounds.  Hence the highest flux models
would seem excluded.  However, the uncertainty of the predicted
antiproton fluxes are about a factor of 5 and our estimate of the
increased antiproton flux is uncertain by at least a factor of 2. 
Hence, even the highest flux models may be marginally consistent with the
antiproton limits from BESS. For this reason we have chosen not to 
exclude them from our plot. We also note that even if we pick a model
a factor of 10 lower, we would still get a continuous gamma ray flux
significantly higher than the background, without having any problems
with the antiproton fluxes.

\subsection{Uncertainties}
We end this section with a short discussion of the uncertainties.  As
mentioned, the value of $J(\eta)$ depends strongly on the assumed density
profile for the clumps themselves.  Here, only the
self-interaction cut-off has been applied, which means that these
predictions should be regarded as rather optimistic.  On the other
hand, $J(\eta)$ is averaged over a quite large solid angle,
$\Delta\Omega=1^{\circ}\!\times1^{\circ} \simeq 0.13$ sr, whereas e.g.\
GLAST will have an angular resolution of about $\Delta\Omega=10^{-5}$
sr.  Due to the effect of individual clumps, if one would bin the sky
in $\Delta\Omega=10^{-5}$ sr bins, the fluctuations would be much
larger than those seen in Fig.~\ref{fig:moore}, i.e.\ the clumps would 
appear as hot spots on the sky.


\section{Caustic rings of Dark Matter}
\label{sec:caustics}

As another example of a model having halo structure which could give
rise to potentially observable dark matter annihilation signals we now
consider smooth dark matter infall onto a pre-existing galaxy.  We
will employ the model of Sikivie \cite{sikivie992} for the formation
of caustic rings of dark matter.  We first review the parts of
Sikivie's model needed to calculate the gamma ray flux from the
caustics.

\subsection{Infall model}
\label{subs:infallmod}

The dark matter particles, assumed to be collisionless and having a
very low intrinsic velocity dispersion, are initially put on a
spherical shell.  This shell will then oscillate onto and out of the
galaxy producing inner and outer caustics.  If the particles have
initial net angular momentum, caustic rings perpendicular to the axis of
rotation are formed.  The caustic rings will be a persistent
feature in space, as there will always be shells turning around.

By definition, the density is strongly increased where a caustic is
formed.  In fact,
if the particles have vanishing initial velocity dispersion,
$\delta_{\rm v}$, the
density diverges.  We will neglect the velocity dispersion when
deriving the general shape and location of the caustics, but introduce
$\delta_{\rm v}/c=10^{-13}$ 
when we consider the detailed density distribution close to the
caustics.  The value of the velocity dispersion is typical of what is
expected for a WIMP with a mass of the order of 100 GeV. The reason is
that although the relic density of a WIMP of mass $m$ is fixed by the
freeze-out from chemical equilibrium at the high temperature of around
$m/20$, it will stay in kinetic equilibrium through weak interactions
until a temperature $T_W$ around 1 MeV. The primordial velocity
dispersion is thus roughly $\sqrt{3T_W/m}\sim 5\cdot 10^{-3}\sqrt{100\
{\rm GeV}/m}$.  The redshift factor since $T_W\sim 1$ MeV is of the
order of $6 \cdot 10^{11}$, giving the quoted result. 
It is worth to point out, however, that the effective
velocity dispersion due to e.g.\ a clumpy infall might be much higher,
significantly changing our results.  We will come back to this issue
in section \ref{sec:uncert}.

\subsection{The density profile}
\label{densprof}

We label the particles arbitrarily by a
3-parameter, $\vek{\alpha}$ (which could, for instance, be the
position of the particle at a given initial time).  The flow of a
particle is completely specified by giving for each time its spatial
coordinate $\vek{\rm x}(\vek{\alpha},t)$.  If we have $n$ different flows
at $\vek{\rm x}$ and $t$, we can write the solutions of
$\vek{\rm x}=\vek{\rm x}(\vek{\alpha},t)$ as $\vek{\alpha}_{j}(\vek{\rm x},t)$,
where $j=1,\ldots ,n$.  To obtain the total number of particles, $N$,
we integrate the number density of particles over
$\vek{\alpha}$-space,
\begin{equation}
\label{eq:totnumpart}
N=\int\frac{d^3N\left(\vek{\alpha}\right)}{d\alpha_{1}d\alpha_{2}d\alpha_{3}}
d^{3}\alpha.
\end{equation}
Mapping onto position space gives the number density
\begin{equation}
\label{eq:numdens}
  d\left(\vek{\rm x},t\right)=\sum_{j=1}^{n}\frac{d^3N\left(\vek{\alpha}_{j}\left(\vek{\rm x},t\right)\right)}
{d\alpha_{1}d\alpha_{2}d\alpha_{3}}
\frac{1}{\left|D\left(\vek{\alpha},t\right)
  \right|_{\mbox{\scriptsize{$\vek{\alpha}_{j}\left(\vek{\rm x},t\right)$}}}}.
\end{equation}
where 
${\rm det}\left(\frac{\partial{\mbox{\footnotesize{\vek{\rm x}}}}}
{\partial{\mbox{\footnotesize{\vek{\alpha}}}}}\right)\equiv D\left(\vek{\alpha},t\right)$
is the Jacobian of the map $\vek{\alpha}\rightarrow \vek{\rm x}$.
Wherever $D\left(\vek{\alpha},t\right)=0$, the density will diverge, and hence caustic surfaces
are associated with zeros of $D$.

\begin{figure}
\centerline{\epsfig{file=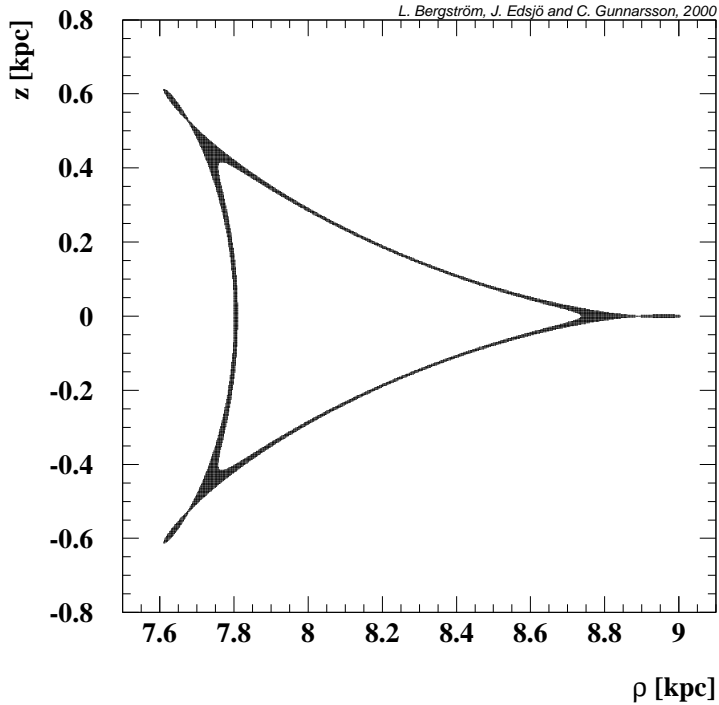,width=0.6\textwidth}}
\caption{
  Plot of points where the density exceeds 1.0 GeV/cm$^{3}$. Note that
  for ${\cal R}_{0}=$7.9, 8.2 and 8.5 kpc, we are situated inside
  the tricusp.}
\label{fig:tricuspquant}
\end{figure}

We assume that the flow of particles is axially symmetric
about the $\hat{z}$-axis (coinciding with the rotation axis of the Galaxy)
and also reflection symmetric with
respect to the $\hat{x}$-$\hat{y}$-plane, i.e.~under reflection
$z\rightarrow-z$. We also assume the dimensions of the cross-section
of the caustic ring to be small compared to the ring radius.  
Let $R(t_{0})$ be the turnaround radius for a shell at time $t_{0}$
in the $z=0$
plane and let $a$ be the ring radius. We then parameterize the flow as
$\vek{\rm x}(\theta_{0},\varphi_{0},t_{0};t)$, where $\vek{\rm x}$ is the
position vector at time $t$ of the particle that was at polar and
azimuthal angles $\theta_{0}$ and $\varphi_{0}$ on the sphere of radius
$R(t_{0})$ at $t=t_{0}$. Axial symmetry suggests we use cylindrical
coordinates with
$\varphi$-independence. Therefore we let $\rho(\alpha,t_{0};t)$ and
$z(\alpha,t_{0};t)$ be the cylindrical coordinates at time $t$ of the
\emph{ring} of particles initially (at $t_{0}$) at
$\theta_{0}=\pi/2-\alpha$. The number density can now be shown to be
\cite[Eq.(4.1)]{sikivie992} 
\begin{equation}
  \label{eq:dens}
  d\left(\rho,z,t\right)=\frac{1}{2\pi\rho}\sum_{j=1}^{n}\frac{d^2N\left(\alpha,t_{0}\right)}
{d\alpha dt_{0}}
\frac{1}{\left|D_{2}\left(\alpha,t_{0}\right)\right|}
  \bigg|_{\left(\alpha,t_{0}\right)=\left(\alpha,t_{0}\right)_j},
\end{equation}
where
\begin{equation}
  D_{2}\left(\alpha,t_{0}\right)=\left|\begin{array}{cc}
  \frac{\partial\rho}{\partial\alpha} &
  \frac{\partial\rho}{\partial t_{0}}\\
  &\\
  \frac{\partial z}{\partial\alpha} & \frac{\partial z}{\partial t_{0}}
  \end{array}\right|,
\end{equation}
and $\left(\alpha,t_{0}\right)_{j}$ are the solutions of the equations
\mbox{$\rho=\rho\left(\alpha,t_{0};t\right)$} and \mbox{$z=z\left(\alpha,t_{0};t\right)$}. 

In this case, the caustic condition $D_2=0$ becomes a fourth-degree
equation in $\alpha$ having four solutions, some of which may be
complex and hence unphysical.  A closer inspection shows that a
caustic is a border between regions with different numbers of flows. 
In Fig.~\ref{fig:tricuspquant} we show the cross section of the fifth
caustic ring (which is the one closest to us), where regions with a
density exceeding 1 GeV/cm$^{3}$ have been indicated. 
We see that the caustic ring resembles a `tricusp'.  Inside the
tricusp, there are four flows and outside there are two\footnote{We do
not take into account the flows not associated with the caustic.}. 
This implies that the sum in Eq.~(\ref{eq:dens}) should have two
(four) terms if we are outside (inside) the tricusp, corresponding to
the number of real roots to the caustic condition.

\begin{figure}
\centerline{\epsfig{file=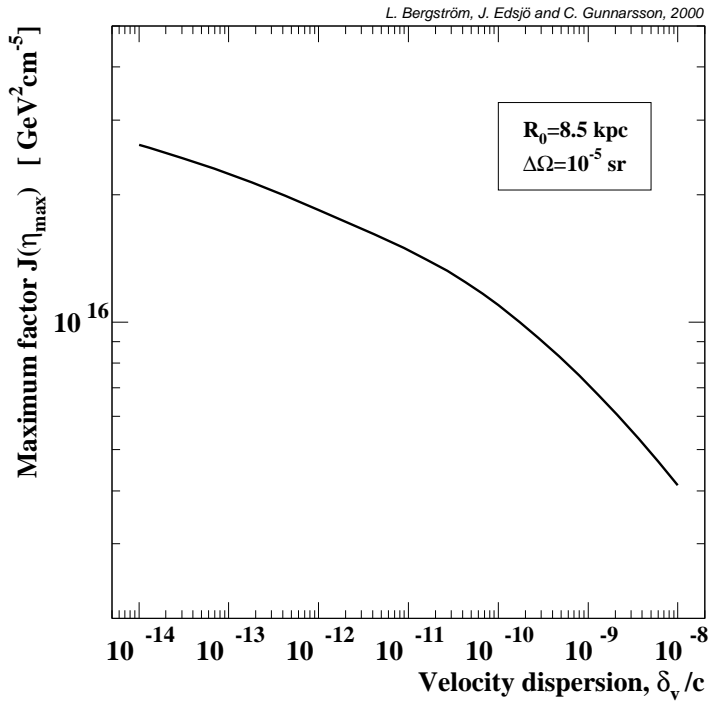,width=0.6\textwidth}}
\caption{The maximum flux, $J(\eta_{\rm max})$ as a function of
the velocity dispersion, $\delta_{\rm v}/c$. As the velocity dispersion
increases from about $10^{-14}$ to $10^{-8}$, the cutoff density decreases
about three orders of magnitude, from 2500 GeV/cm$^3$ to 2.5 GeV/cm$^3$ (with
800 GeV/cm$^3$ at $\delta_{\rm v}/c=10^{-13}$).}
\label{fig:jmvsvd}
\end{figure}

The model of Sikivie has a set of caustic parameters that
describe it.
To find these we assume that the turnaround sphere
is initially rigidly rotating and that it initially really is a sphere, not
just an axially symmetric topological sphere. As mentioned, the axis of 
rotation for the sphere is assumed to coincide with the axis of rotation of the
luminous parts of the Galaxy. We also have to make an assumption about 
the distribution of the smooth component of the dark matter 
distribution (i.e.~not associated with the caustic flows). We adopt
the time-independent potential
\begin{equation}
\label{eq:potmod}
U(r)=-v_{\rm rot}^{2}\ln \left(\frac{R}{r}\right),
\end{equation}
since this potential produces perfectly flat rotation curves with
rotation velocity $v_{\rm rot}$. For comparison, we have also used
the modified isothermal sphere with a density distribution
\begin{equation}
\label{eq:isodens}
  {\cal D}_{\rm iso}(r)={\cal D}_{0}\frac{a_{\rm c}^{2}+
  {\cal R}_{0}^{2}}{a_{\rm c}^{2}+r^{2}},
\end{equation}
where ${\cal D}_0$ is the local dark matter density, ${\cal R}_0$ is
our galactocentric distance and $a_{\rm c}$ is the the scale radius, without
obtaining any significant changes.  To obtain the caustic parameters,
we have followed the procedure in \cite{sikivie992}.  The interested
reader can find the result and more details about the caustic
parameters in Ref.\ \cite{cgexjobb}.  We do not give them here since
the actual values themselves are not very illuminating.

To find the density profile we first rewrite Eq.~(\ref{eq:dens}) into
a more useful form. By reparameterizing the equation according to 
$t_{0}=\tau -\Delta t_{0}(\alpha)$ we have
$dt_{0}=d\tau$.  Due to axial symmetry, the solid angle $d\Omega=\sin
\theta d\theta d\varphi$ can be rewritten as $d\Omega=2\pi\cos \alpha
d\alpha$.  Noting that $dN=dM/m_{\chi}$ with M being the total mass of
particles with mass $m_{\chi}$ we get
\begin{equation}
\label{eq:prefac}
\frac{d^{2}N}{d\alpha dt_{0}}=\frac{2\pi \cos
  \alpha}{m_{\chi}}\frac{d^{2}M}{d\Omega d\tau}. 
\end{equation}
As mentioned earlier, in this model the Solar system should be 
closest to caustic ring number five, and 
following Ref.~\cite{sikivie982}, we find, for this caustic ring,
\begin{equation}
\label{eq:prefacnum}
\frac{d^{2}M}{d\Omega
  d\tau}=10^{-2}\frac{V(0)v_{\rm rot}^{2}}{2\pi G},
\end{equation}
where $G$ is Newton's gravitational constant and $V(0)$ is the 
velocity at the point of closest approach to the Galactic center, i.e.~at 
the caustic.
This was obtained via a self-similar infall model using a scale
parameter $\epsilon=0.2$ defined in Ref.~\cite{sikivie971}. However, the model 
dependence on $\epsilon$ is quite
weak \cite{sikiviepriv}.

To obtain the mass density from Eq.~(\ref{eq:dens}) we must multiply
$d(\rho,z,t)$ by $m_{\chi}$, which cancels the factor $1/m_{\chi}$ in
Eq.~(\ref{eq:prefac}).  Finally, by combining Eqs.~(\ref{eq:dens}),
(\ref{eq:prefac}) and (\ref{eq:prefacnum}), we can obtain the value
for the mass density, ${\cal D}$, of dark matter close to the fifth
caustic ring in the Milky Way.

A diverging density at the caustics results from our assumption of
zero velocity dispersion, which of course is an over-simplified
assumption.  We thus reintroduce a non-zero velocity dispersion by
estimating how much a given velocity dispersion would smear the
caustic.  We do that by considering a particle falling into the
potential $U(r)$.  If we change the initial velocity of the particle
with the velocity dispersion, we obtain a difference in the location
of the point of closest approach (i.e.~the location of the caustic
ring).  We can then use this difference as an estimate of how much the
caustic ring is smeared by the velocity dispersion.  The simplest way
to take the smearing into account is to apply a cut-off in the density
whenever we are closer to the caustic than the smearing scale.  For a
velocity dispersion of $\delta_{v}/c=10^{-13}$, this corresponds to a
cut-off in the density at ${\cal D}_{\rm cut} \simeq 800$
GeV/cm$^{3}$.  Since the density only diverges as $\sim 1/\sqrt{l}$
with $l$ being the distance to the caustic \cite{sikivie981}, we are
not very sensitive to the actual value of the cut-off density as can
be seen in Fig.~\ref{fig:jmvsvd}.  In our calculations we have used a
cut-off density of ${\cal D}_{\rm cut}=800$ GeV/cm$^{3}$.

For comparison we have also used a gaussian smearing of the density
distribution with the same smearing length scale as the cut-off length
scale.  The two methods give practically the same result and we have
used the cut-off method in our actual calculations.

\begin{table}
 \caption{Summary of the peak values of $F_{\Delta\Omega}(\eta)$  for
 the cases considered.}
 \label{tab:jetas}
 \begin{tabular}{llll} 
 $\Delta\Omega$ & ${\cal R}_{0}$ & $\eta_{\rm max}$ &
 $F_{\Delta\Omega}(\eta_{\rm max})$ \\
 {}[sr] & [kpc] & [rad] &
 [10$^{21}$GeV$^{2}$cm$^{-5}$sr$^{-1}$]  \\ \hline
 10$^{-1}$ & 8.5 & 0.54 & 0.26 \\
  & 8.2 & 0.78 & 0.22 \\
  & 7.9 & 1.15 & 0.29 \\\hline
 10$^{-3}$ & 8.5 & 0.54 & 0.96 \\
  & 8.2 & 0.79 &  0.91 \\
  & 7.9 & 1.17 & 1.34 \\\hline
 10$^{-5}$ & 8.5 & 0.53 & 2.20 \\
  & 8.2 & 0.79 & 4.05 \\
  & 7.9 & 1.18 & 3.05 \\
 \end{tabular}
\end{table}

In Fig.~\ref{fig:tricuspquant}, all points where
${\cal D}(\rho,z)>1.0$ GeV/cm$^{3}$ are plotted.  From the figure
we note that for realistic values of ${\cal R}_{0}$ of about 8--8.5
kpc, we are located \emph{inside} the tricusp, which is very
interesting from the point of view of the possibility of detection.

Now focus on $F_{\Delta\Omega}(\eta)$.  For ${\cal R}_{0}=$7.9, 8.2
and 8.5 kpc the angular range $0<\eta\lesssim\pi$ was scanned for
three different typical angular acceptances, $\Delta\Omega=10^{-1},
10^{-3}$ and $10^{-5}$ sr.  All these scans have maxima at the angles
corresponding to the cusps, since the density is strongly enhanced
there.  The symmetry implies a peak also at $-\eta$ if there is a peak
at $\eta$.  The maximum value of $F_{\Delta\Omega}(\eta)$ from all
scans was found at $\eta\simeq 0.79$
rad, $\Delta\Omega=10^{-5}$ sr and ${\cal R}_{0}=8.2$ kpc.  Table
\ref{tab:jetas} gives the maximum values of $F_{\Delta\Omega}(\eta)$
for the different $\Delta\Omega$ and ${\cal R}_{0}$.  The angle which
gives the maximum of $F_{\Delta\Omega}(\eta)$ is denoted $\eta_{\rm max}$.

\subsection{Background to signal comparison}

\begin{figure}
\centerline{\epsfig{file=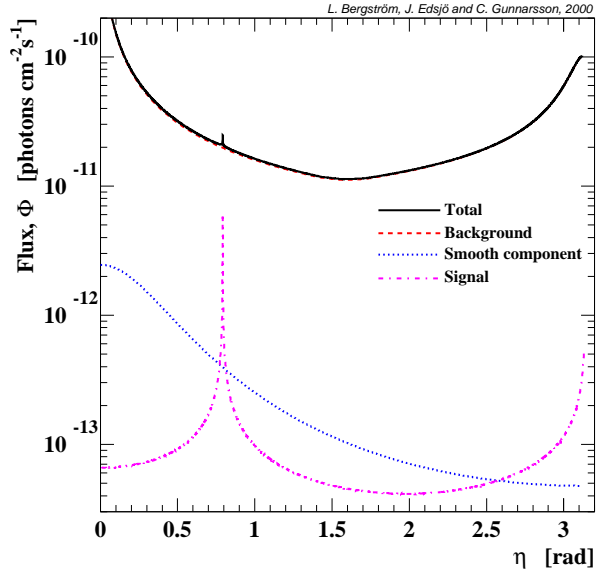,width=0.5\textwidth}}
\caption{Flux of gamma rays above 1 GeV from the caustic, the smooth
halo, the background and the sum of them for $\Delta\Omega=10^{-5}$
sr.  A maximum flux SUSY model was used for the signal and the smooth
part.  For the smooth halo an isothermal sphere with the scale radius
$a_{\rm c}=4$ kpc and the local density ${\cal D}_{0}=0.2$ GeV/cm$^{3}$
was used.  Our galactocentric distance was set to ${\cal R}_{0}=8.2$
kpc, but the results are essentially the same for other values.}
\label{fig:total}
\end{figure}  

To see whether the signal is potentially detectable, we have plotted
the signal of continuous gamma rays from the caustic ring, the flux
from annihilations in the smooth halo of Eq.~(\ref{eq:isodens}), the
background and the sum of the three in Fig.~\ref{fig:total} for
$\Delta\Omega=10^{-5}$ sr.  For the smooth part we used ${\cal
D}_0=0.2$ GeV/cm$^3$ and $a_{\rm c}=4$ kpc. (The reason for not putting
${\cal D}_0=0.3$ GeV/cm$^3$ as before is that we expect about 1/3 of
the dark matter to be in the caustic flow.)
Our galactocentric distance was
set to ${\cal R}_{0}=8.2$ kpc, but the results are essentially the
same for other values.  The SUSY model used for the signal and smooth
parts is one of maximum ${\cal S}$ in Fig.~\ref{fig:sfactors}.  The
figure shows that the signal is quite small even for such a SUSY model
and as is implied in Fig.~\ref{fig:sfactors}, most models produce a
flux several orders of magnitude smaller which would make the
signal vanishingly small compared to the background.  Hence, the flux
shown in Fig.~\ref{fig:total} should be regarded as a best-case scenario
with the highest possible flux from the caustic rings.  We thus
conclude that the possibility of detection in the caustic ring model,
unlike the hierarchical clustering model, is quite marginal.  In
Fig.~\ref{fig:total}, we plotted the continuous gamma ray flux, and
for the gamma ray lines, the figure would look essentially the same,
but with fluxes about a factor of $5 \times 10^{-5}$ lower.

\subsection{Intensity pattern on the sky}

\begin{figure}
\centerline{\epsfig{file=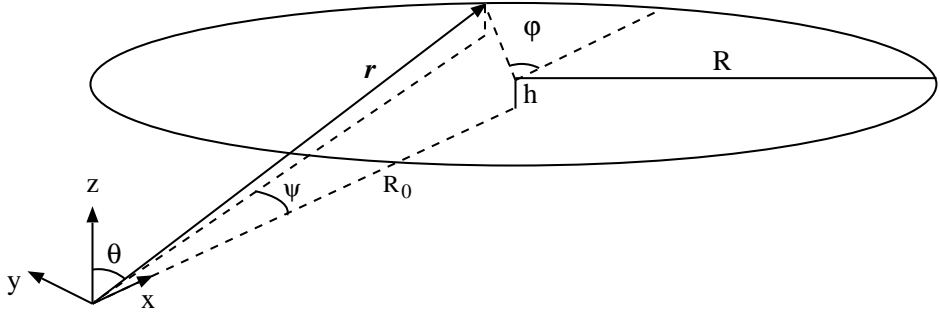,width=0.7\textwidth}}
\caption{Definition of the coordinate system used.}
\label{fig:ring}
\end{figure}

The signal from the caustic ring is very narrowly
localized in the sky angle $\eta$.  To investigate how the full
signal pattern would appear on the sky, we can then make the simplified assumption
that the source function for the gamma-ray emission is given by a
delta-function $\propto\delta^{(3)}\left(\vek{\rm r}
-\vek{\rm r}_0(\varphi)\right)$, where $\varphi$ is an azimuthal angle
parameterizing the ring (see Fig.~\ref{fig:ring} for the notation).

Since each mass element of the ring locally gives rise to the same
isotropic gamma-ray flux, the signal seen near the Earth's location
can be estimated analytically by geometrical considerations.
Define spherical coordinates on the celestial sky $\theta$ and $\psi$ 
where $\theta$ is the polar angle (measured with respect to a
z-axis which is tied to the Solar system and pointing perpendicularly
to the plane of the Galaxy), and $\psi$ is an azimuthal angle with
$\psi=0$ corresponding to the direction of the Galactic center.

We now consider one of the two closest rings corresponding to the cusps 
at $z\neq 0$ in Fig.~\ref{fig:tricuspquant}, which are inside our position in
the Galaxy, let us take the one which has $z=h>0$, and radius $R$. 
(Due to the $z$ symmetry, 
the two rings give precisely
the same signal.) The location vector on the ring can be written
\begin{equation}
\vek{\rm r} = (R_0,0,h)+R(\cos\varphi,\sin\varphi,0),
\label{eq:1}
\end{equation}
where the point nearest to us has $\varphi=\pm\pi$, and the furthest
point corresponds to $\varphi=0$.
Both angles $\theta$ and $\psi$ can now be computed:
\begin{equation}
\cos\theta={h\over |\vek{\rm r}|}
\label{eq:2}
\end{equation}
and
\begin{equation}
\tan\psi={R\sin\varphi\over R_0+R\cos\varphi}.
\label{eq:3}
\end{equation}
Ring elements of constant flux $Rd\varphi\vek{\rm e}_\varphi$ are now mapped onto
the sky as
\begin{equation}
Rd\varphi\vek{\rm e}_\varphi \rightarrow |\vek{\rm r}|d\theta \vek{\rm e}_\theta+
|\vek{\rm r}|\sin\theta d\psi \vek{\rm e}_\psi 
 = |\vek{\rm r}|\left({d\theta\over d\varphi}\right)d\varphi \vek{\rm e}_\theta+
|\vek{\rm r}|\sin\theta \left({d\psi\over d\varphi}\right)d\varphi \vek{\rm e}_\psi.
\label{eq:4}  
\end{equation}
From this we can read off the inverse of the Jacobian, which is the
``magnification'' $M$,
\begin{equation}
M={R\over |\vek{\rm r}|^3\sqrt{\sin^2\theta\left({d\psi\over d\varphi}\right)^2+
\left({d\theta\over d\varphi}\right)^2}}.
\label{eq:5}
\end{equation}
where an extra factor of $|\vek{\rm r}|^2$ in the denominator
accounts for the geometrical
fall-off of the flux with the square of the distance.
Introducing the dimensionless parameters
\begin{equation}
\kappa={h\over R} \qquad ; \qquad
\rho={|\vek{\rm r}|\over R} \qquad ; \qquad
\xi={R_0\over R},
\end{equation}
we obtain, after some algebra,
\begin{equation}
\left({d\theta\over d\varphi}\right)^2={\kappa^2\over \rho^2-\kappa^2}{\xi^2\over \rho^4}\sin^2\varphi,
\label{eq:7}
\end{equation}
and
\begin{equation}
\left({d\psi\over d\varphi}\right)^2={\left(1+\xi\cos\varphi\right)^2\over \left(1+\xi^2+2\xi\cos\varphi\right)^2},
\label{eq:8}
\end{equation}
from which we can compute the flux as a function of angle on the sky by use
of Eq.~(\ref{eq:5}).

\begin{figure}
\centerline{\epsfig{file=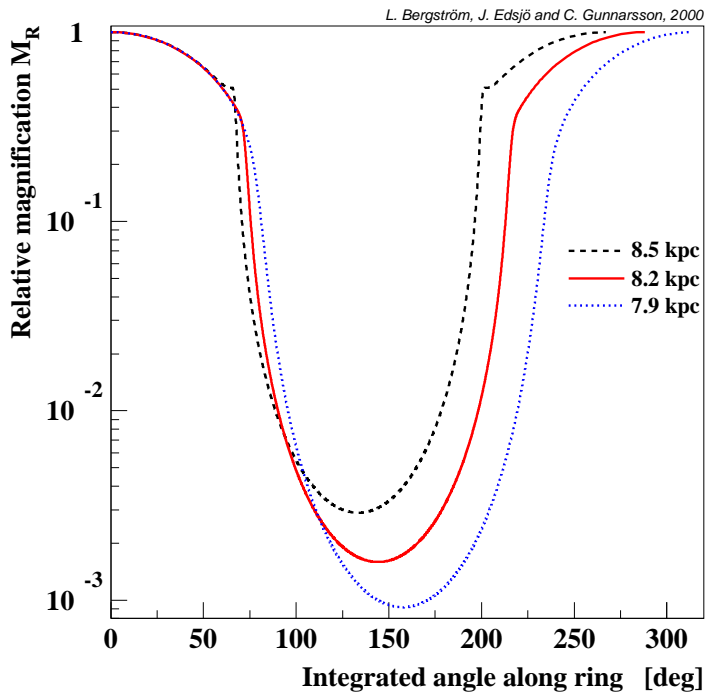,width=0.6\textwidth}}
\caption{Relative magnification $M_R$ as a function of
  the integrated sky angle along the ring for three different galactocentric
  distances, ${\cal R}_0$.}
\label{fig:ring2}
\end{figure}

We now normalize the magnification to 1 for the part of the ring that 
is closest to us. In Fig.~\ref{fig:ring2} we show the magnification 
as a function of the integrated space angle along the caustic ring. We clearly 
see that the signal is fairly high out to about 60$^{\circ}$ from 
the direction to the galactic center. Further away, the flux drops 
dramatically because of the geometric fall-off with distance.

\subsection{Detection potential}

We will now investigate if the Gamma-ray Large Area Space Telescope
(GLAST) \cite{glast} would be able to see even the best-case signal.

We assume the same maximal SUSY model as in Fig.~\ref{fig:total} and
use $\Delta\Omega=10^{-5}$ sr, which is close to GLAST's expected
angular resolution.  We then integrate the signal over the nearest
100$^{\circ}$ of the caustic ring, where the average magnification
(compared to the closest part of the ring) is around 0.8 according to
Fig.~\ref{fig:ring2}.  We further assume that the average effective
area will be $\langle A_{\rm eff} \rangle = 5000$ cm$^{2}$ and that we
integrate for one year.  In this strip we would then expect around 400
events of continuous gamma rays above 1 GeV from the caustic ring and
about 1700 events from the diffuse background as measured by EGRET.
Hence, this would not be a prominent signature on the sky, especially
since generic MSSM neutralinos give much lower rates.  The total
number of events from the gamma ray lines would be about 0.01 even in
this optimistic scenario, so the prospects of detecting these are
essentially zero. We end by noting that the uncertainty in the 
background estimate is at least a factor of two, but this hardly 
changes our conclusions that the gamma ray signal from the caustics 
is very hard to detect.

Just like in the scenario with clumps, ACTs could be used as a
follow-up \emph{if} GLAST would see an indication of a caustic.

\subsection{Antiprotons}

To estimate the increase in the antiproton flux from the caustics, we
have integrated ${\cal D}^{2}$ with a cut-off density of 800 GeV/cm$^3$ over
the region $z \in [-0.7,0.7]$ kpc and $\rho \in [7.5,9.1]$ kpc and we
find that the total annihilation rate from the caustic is about 3
times higher than that from a smooth halo in the same region.  Since
the antiproton flux depends mainly on the total annihilation rate
within the closest few kpc we don't expect the flux of antiprotons to
increase by more than a factor of 3 in the caustic scenario.  We then
see that for the highest values of ${\cal S}_{\gamma~\!{\rm cont}}$, we
would get an antiproton flux that is higher than the BESS measurements
by a factor of about 3.  However, the antiproton flux prediction can
be uncertain by as much as a factor of 5, so it is still possible that
these models with high ${\cal S}_{\gamma~\!{\rm{cont}}}$ are consistent
with the antiproton measurements and we have thus chosen not to
exclude them.

\subsection{Uncertainties}
\label{sec:uncert}

We have two classes of uncertainties in this derivation of the gamma
ray flux from the caustic rings.  The first one comes from the fact
that we don't know the MSSM parameters and this alone gives an
uncertainty of several orders of magnitude, as seen in
Fig.~\ref{fig:sfactors}.  The second one are the uncertainties in the
caustic model by Sikivie. 

The main uncertainty is the assumption of smooth continuous infall of
collisionless dark matter with a very small velocity dispersion, of
the order of $\delta_{\rm{v}}/c=10^{-13}$.  As $N$-body simulations
seem to suggest, structure forms hierarchically and the infall to our
galaxy should not be smooth, but rather clumpy.  In this case, we
would get an effective velocity dispersion much higher than
$10^{-13}$.  Some of the structures of the caustics might remain, but
the significant density increase we have found here would be washed
out and the signal could be reduced by orders of magnitude.  On the
other hand, in this case, the signal from the clumps themselves could
be detectable \cite{moore} as we saw in Section~\ref{sec:nbody}.

The infall model itself also has some uncertainties.  For instance, we
have assumed that the infalling sphere is rigidly rotating, which
might not be a realistic approximation.  We have also assumed that the
axis of rotation is the same as that of the luminous matter in our
galaxy.  This might depend on the details of how the bulge and disk
were formed, and need not be the case.

Given these uncertainties, our strategy has been to investigate if 
there is a detectable signal even with the most optimistic assumptions. 
We have found that the detection potential is \emph{very} weak although 
not zero under extremely optimistic assumptions.

\section{Conclusions}
\label{sec:conclusions}

We have supplemented the recent work by Calc\'aneo-Rold\'an and
Moore based on $N$-body simulations of structure formation in
cold dark matter models, by giving absolute rate predictions 
in the MSSM for the
gamma-ray signal expected by the clumpy substructure of these
simulated halos.  The predicted rates are quite high, making this
a promising signal to search for, both concerning continuous gamma-rays
and, if supersymmetric parameters are favourable, the distinctive
monoenergetic gamma-ray lines predicted if the dark matter indeed
consists of WIMPs. In particular, the upcoming GLAST space-borne
gamma-ray detector will be an ideal instrument searching for 
these intriguing patterns on the sky.

Motivated by the work done by Pierre Sikivie on caustic rings of dark
matter, we have also estimated the gamma ray flux from these.
However, even with very optimistic assumptions about
the infall model and velocity dispersion, we can get a signal of 
continuous gamma rays that is only
marginally detectable by GLAST. The uncertainties in the Sikivie model
are large and relaxing some of the assumptions could reduce the
flux further by several orders of magnitude.

However, it is worth stressing that if we relax the assumption 
of a smooth continuous infall, we would reduce the flux from the 
caustics drastically, but we would at the same time enhance the flux 
from the infalling clumps.

We have also investigated how much the antiproton flux is expected to 
increase in these two scenarios and found that for the models with the 
highest flux of continuous gamma rays, we would violate the BESS bound 
on antiprotons by a factor of about 3 in the caustic scenario and 
5--10 in the hierarchical clustering scenario. Taking the 
uncertainties of the antiproton prediction into account, this would at 
best be marginally allowed by the BESS measurements. For the gamma ray 
lines, the correlation with the antiproton flux is weaker, and we can 
easily find models with high fluxes of monochromatic gamma rays that 
do not violate the BESS bounds.

We end by concluding that it is interesting that in two such
orthogonal scenarios of galaxy formation, the outcome in both may be
the existence of dark matter density enhancements which may give
observable signals in upcoming gamma-ray detectors.  This may indicate
that the possibility of detection will exist also for refined models
which describe the Milky Way dark matter distribution more
realistically.

\begin{acknowledgments}

L.B., J.E.\ and C.G.\ wish to thank the Swedish Natural Science
research Council (NFR) for support and L.B.\ and J.E.\ also thank  
the Aspen Center for Theoretical Physics, where parts of this 
work was done, for hospitality. We all want to thank Pierre Sikivie,
Piero Ullio and Larry
Widrow for useful discussions,
and  Carlos Calc\'aneo-Rold\'an and Ben Moore for providing their
simulation results in numerical form and for comments.

\end{acknowledgments}


\end{document}